\documentclass[aps,pre,twocolumn]{revtex4-1}
\usepackage{amsmath}
\usepackage{graphicx}
\usepackage{float} 
\usepackage{amssymb}
\usepackage[section]{placeins}

\usepackage{comment}
\begin{document}
\title{Precursor Solitons in Plasma Flow Past Charged Obstacles: Role of Obstacle Bias and Ion Temperature Anisotropy}
\author{Prince Kumar$^{1}$, S. K. Mishra$^{1}$} 
\affiliation{$^{1}$Planetary Sciences Division, Physical Research Laboratory, Ahmedabad, Gujarat 380009, India }
\date{\today}
\begin{abstract}
We investigate electrostatic ion-acoustic precursor solitons in a plasma flow past an absorbing charged obstacle using two-dimensional (2D) electrostatic PIC simulations. 
A key outcome of the present formulation is that ion-temperature anisotropy can enable precursor formation even in regimes where isotropic plasmas, due to Landau damping, cannot sustain such structures. 
Specifically, temperature anisotropy in the 2D drifting flow along the \(x\)-direction, arising from a reduction in the transverse thermal velocity (\(y\)-direction) relative to the parallel thermal velocity (\(x\)-direction), favors the generation of coherent upstream structures whose effectiveness increases with stronger anisotropy. 
%
%
%
Both positive and negative obstacle polarities are considered to identify the conditions for upstream nonlinear structure formation. 
%
A negatively biased plate produces only a wake-like response, whereas a positively biased plate generates upstream density pulses. 
%
%
%
%
%
%
This study offers physical insight into nonlinear wave formation in streaming plasmas over charged objects and could be useful for plasma-based debris detection in the low Earth orbit (LEO) region.
%
\end{abstract}
\maketitle
\section{Introduction}
\label{sec:introduction}
The low-Earth-orbit (LEO) environment is increasingly crowded with satellites, rocket bodies, and small debris fragments~ \cite{leal2025managing,ocal2024inevitable,reynaud2026space}. Even subcentimeter particles pose a serious hazard because they travel at high velocity and can cause severe impact damage. Since these small objects are difficult to detect with conventional radar or optical systems, novel in situ methods are needed for debris detection and collision avoidance~\cite{mcdonnell1993near,christiansen2004space}. One promising idea is to use the ambient plasma as a diagnostic medium ~\cite{sen2015,bernhardt2023observations,hughes2025observations}. As a fast-moving object traverses through the ionospheric plasma, it becomes charged and disturbs the surrounding medium, potentially forming nonlinear structures. Among these, precursor solitons are especially attractive because they propagate ahead of the obstacle and may serve as early signatures of a moving debris object. The possibility of debris-generated precursor solitons was first proposed by Sen and collaborators using a weakly nonlinear fluid model~\cite{sen2015}. 
Later studies extended this framework to multidimensional geometries and LEO-relevant plasma conditions, examining the properties of the resulting precursor structures~\cite{truitt2021three, Truitt2020}. Laboratory experiments in flowing dusty plasmas have verified the generation of precursor solitons by charged obstacles~\cite{arora2021experimental, bandyopadhyay2022driven,jaswal2016}. The effect of magnetic field, obstacle geometry, and different flow-angle configurations on precursor generation and their sustenance against Landau damping has been demonstrated in ~\cite{dharodi2026magnetic,ganguli2025orbital,kumar2020precursor,sen2023electromagnetic}. Together, these theoretical, numerical, and experimental works establish that charged orbital debris could produce detectable nonlinear plasma signatures in the ionosphere.

However, a key difficulty arises in the real ionosphere: electrostatic ion-acoustic waves are subject to Landau damping unless the ion thermal speed is much smaller compared to the wave phase speed. In practice, this requires the electron temperature to exceed the ion temperature for the waves to propagate longer without significant damping. 
%
First-principles kinetic simulations have shown that under typical LEO plasma conditions (where $T_i\sim 0.1~T_e$ or higher), coherent precursor structures do not form due to Landau damping and other effects~\cite{resendiz2024ion,dharodi2026plasma, sam2025self,dharodi2026magnetic,sam2026ion}.
%
For example, electrostatic particle-in-cell (PIC) simulations~\cite{resendiz2024ion} have shown that the anticipated upstream soliton train can be significantly modified or suppressed when full kinetic effects of the ions are included. This highlights that kinetic physics are essential for soliton formation and must be properly accounted for in modeling efforts aimed at making realistic, quantitative predictions for soliton-related phenomena.
%
%
These findings pose a challenge for debris detection via plasma waves in electrostatic conditions, as the ion Landau damping washes out the precursor in the ambient plasma, and the excitations predicted by earlier models may be too weak to observe. 

 One factor that has received comparatively little attention is the anisotropy of the ion velocity distribution function. In many space plasma conditions, particularly in the ionosphere, ion temperatures frequently become anisotropic due to mechanisms such as frictional heating, field-aligned acceleration, and other anisotropic energy sources \cite{lockwood1993eiscat,lomidze2021estimation,balmforth1999localized}. In fact, studies have examined subsonic ion flow past an absorbing obstacle with moderate anisotropy ($T_{i\parallel} = 5T_{i\perp}$) and observed the formation of filamentary vortices in the wake region. However, their focus was on downstream wake dynamics rather than on the generation of upstream nonlinear structures.~\cite{guio2004phase} 
%
%
  %
%
Motivated by this fact, the present work investigates whether an anisotropic ion distribution can restore precursor soliton formation even in the Landau-damped regime. Therefore, a 2D setup is prepared and simulated using a 2D electrostatic PIC code that closely aligns with the electrostatic ionospheric parameter space. Using 2D electrostatic PIC simulations of plasma flow over a fixed biased absorbing obstacle, the ion thermal conditions are systematically varied to investigate following cases: (i) a cold-isotropic case ($T_i/T_e\ll1$) as a base case, (ii) a warm isotropic case ($T_i/T_e \sim 0.1$) where Landau damping dominates, and (iii)  the warm case with reduced transverse spread, i.e., $T_{i\perp} < T_{i\parallel}$, where $T_{i\parallel}$ and $T_{i\perp}$ denote the ion temperatures parallel and perpendicular to the drifting plasma direction, respectively.
We consider both positive and negative obstacle biases, since debris surfaces in space plasmas can attain either polarity depending on the relative balance between ambient plasma charging and photoelectron emission under different environmental conditions \cite{resendiz2024ion}.  

Our simulations reveal a clear hierarchy. With a positively biased plate and cold ions ($T_i/T_e = 10^{-3}$), the incoming flow remains coherent and a strong upstream density pulse forms quickly, splitting into a train of solitary-like peaks at later times.  This confirms the classic fluid model result in its most favorable scenario. However, when the ion temperature is raised to $T_i/T_e=0.1$, the same obstacle produces only a broad, diffuse perturbation with no distinct solitons.  The enhanced thermal spread leads to rapid phase mixing and Landau damping of the disturbance, suppressing the nonlinear soliton balance.  In this regime, we are consistent with prior PIC studies: no upstream soliton emerges \cite{resendiz2024ion}. 
Crucially, we then find that introducing ion-temperature anisotropy can reverse this outcome. For flows with reduced transverse temperature (e.g.,\ $T_{i\perp}/T_{i\parallel}=0.1$ or smaller), the ions remain more stream-aligned.  In such cases the obstacle-driven compression stays relatively more localized and coherent. Our results suggest, a significant anisotropy ($T_{i\perp} \ll T_{i\parallel}$) might induce the growth of upstream density peaks resembling solitons even for high temperature ions ($T_i/T_e=0.1$). The stronger anisotropy results in more pronounced soliton structures. 

In contrast, a negatively biased plate produces no upstream solitary pulses; instead, only downstream wake perturbations are observed. This is expected, since a negative obstacle attracts ions, and the upstream compression mechanism is absent. The absence of precursor waves for negative bias is consistent with prior theory \cite{resendiz2024ion,sam2025self} and the simulations presented herein, confirming that the effect is controlled by the nature of the electrostatic forcing and the obstacle boundary conditions. 
%
%
%
The results suggest that the anisotropy widens the parameter space in which debris-induced solitons can survive damping. 
This work provides a framework for assessing precursor visibility under more realistic conditions and motivates the inclusion of anisotropic kinetics in future debris-plasma interaction models. 

The rest of this paper is organized as follows.  In Section~\ref{sec:setup}, we describe the simulation model, normalization, and parameters chosen.  Section~\ref{sec:numerical_results} presents the key results, comparing the isotropic cold case, the warmer isotropic case, and the anisotropic cases for a positively biased obstacle, followed by the negatively biased case.  
Finally, Section~\ref{sec:results_summary} presents the main conclusions and discusses future prospects of the analysis. \\ \\ \\
%
\section{Simulation Setup}
\label{sec:setup}
%
%
We perform 2D electrostatic PIC simulations of a drifting plasma interacting with a fixed biased absorbing plate. Ions are treated kinetically, while electrons are modelled as a Boltzmann fluid, yielding a self-consistent solution of the nonlinear Poisson equation. The schematic of the simulation domain is presented in Fig.~\ref{fig:Schematic}. The technical numerical details are presented in the appendix~\ref{app:numerical details}. The table \ref{tab:sim_params} presents the simulation parameters and their values.
\begin{figure}[H]
\centering
\includegraphics[width=1.0\linewidth, height=6.0cm]{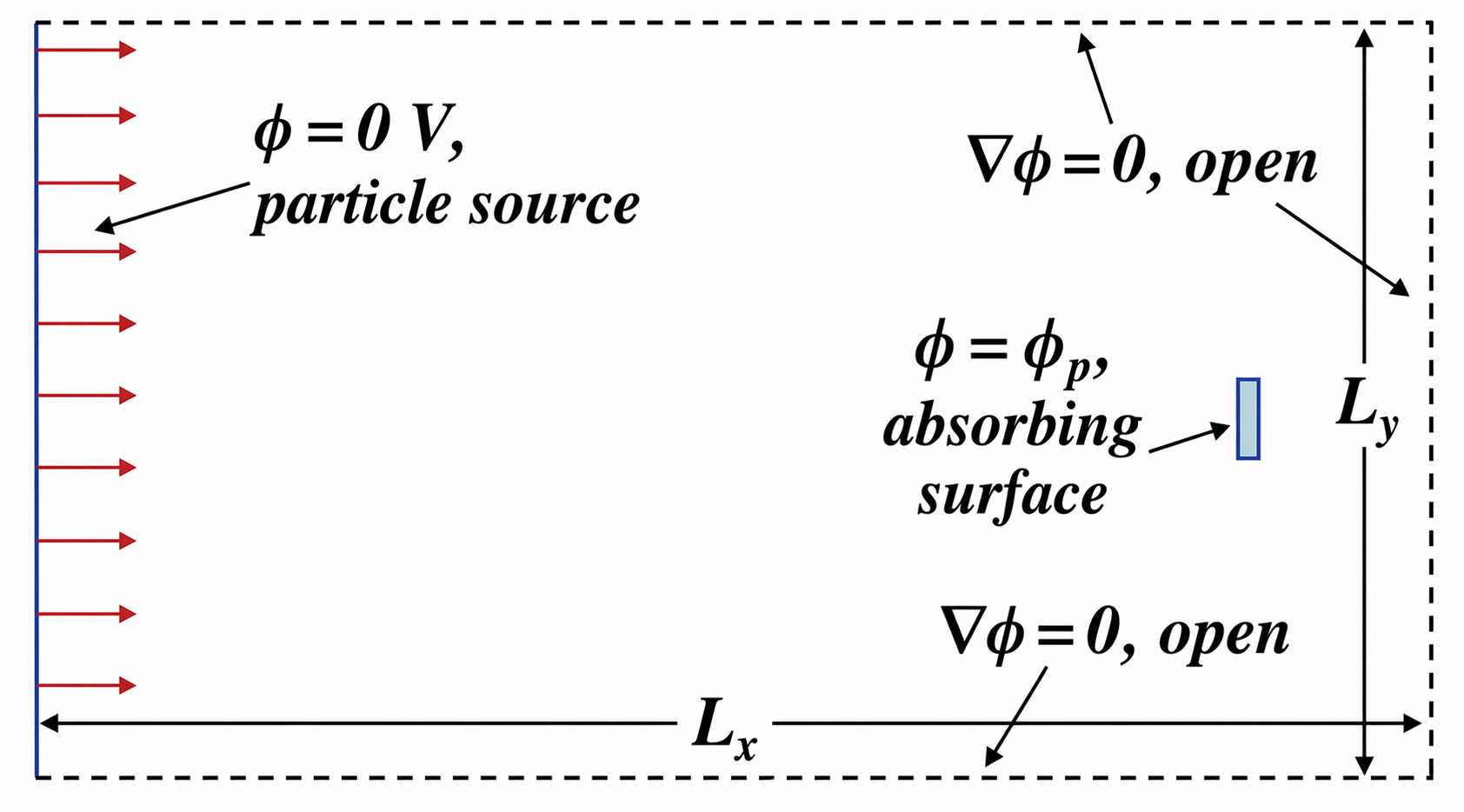}
\caption{Schematic illustration of the two-dimensional rectangular computational domain of dimensions $L_x \times L_y$. The left boundary is maintained at $\phi = 0~\mathrm{V}$ and acts as the particle source. The top, right, and bottom boundaries satisfy open boundary conditions $(\nabla \phi = 0)$. A small obstacle, positioned near the right boundary, is held at potential $\phi = \phi_p$ and represents the absorbing surface.}
\label{fig:Schematic}
\end{figure}

The purpose of this numerical study is to investigate the formation and evolution of precursor structures ahead of a biased debris-like obstacle under ionospheric plasma conditions. We consider both positively and negatively biased debris and analyze the effects of the ion-to-electron temperature ratio and temperature anisotropy on the generation, amplitude, and long-term evolution of precursor solitons. Smaller values of $T_i/T_e$ correspond to colder ions; this parameter is used to probe the significance of Landau damping and to determine whether ion-temperature anisotropy modifies the formation and propagation of precursor structures. We first present results for the cold-ions, $T_i/T_e = 10^{-3}$, which refers to a favorable condition for precursor formation (as a benchmark test case), followed by the intermediate case $T_i/T_e = 0.1$, where kinetic damping becomes important. After the benchmark test, we introduce ion-temperature anisotropy to further examine whether the structure formation is robust in a non-isotropic medium. Finally, we extend the analysis to negative-biased debris potentials to compare the plasma's response to opposite-polarity perturbations.
\begin{table}[h]
\caption{Simulation parameters for drifting plasma.}
\label{tab:sim_params}
\centering
\small
\setlength{\tabcolsep}{5pt}
\begin{tabular*}{\columnwidth}{@{\extracolsep{\fill}} l c c @{}}
\hline\hline
Parameter & Value & Description \\
\hline
Ion species & $\mathrm{O_2^+}$ & oxygen ion \\
Ion mass $M$ & $5.32\times10^{-26}\,\mathrm{kg}$ & $32\,m_u$ \\
Density $n_0$ & $10^{12}\,\mathrm{m^{-3}}$ & Density \\
Electron temp. $T_e$ & $0.3\,\mathrm{eV}$ & Fixed \\
Ion temp. $T_i$ & $3\times10^{-4}$--$3\times10^{-2}\,\mathrm{eV}$ & Varied \\
Debye length $\lambda_D$ & $4.07\,\mathrm{mm}$ &  Debye length \\
$\omega_{pi}$ & $7.37\times10^{5}\,\mathrm{s^{-1}}$ &  plasma freq. \\
$v_{\mathrm{th},i}$ & $13.5$--$420.7\,\mathrm{m/s}$ & thermal speed \\
$C_s$ & $9.51\times10^{2}\,\mathrm{m/s}$ & Sound speed \\
$v_{\mathrm{drift}}$ & ($1.0-7.0)\times10^{3}\,\mathrm{m/s}$ & Drift flow \\
Mach~Number~$M_a$ & $\approx 1.10$ and higher & $v_{\mathrm{drift}}/C_s$ \\
Plate potential~$\phi_p$ & $+0.5$ to $+3$ V & Positive bias \\
 & $-1.5$ V & Negative bias \\
Plate area & $\sim (1-4)\lambda_D^2$ & -- \\
\hline\hline
\end{tabular*}
\end{table}
\section{Numerical Results}
\label{sec:numerical_results}
In this section, we present the results obtained from the simulations of plasma flow past a biased plate under ionospheric conditions.
%
The electron temperature is fixed at $T_e = 0.3~\mathrm{eV}$, the incoming ion flow speed is $v_d = 1~\mathrm{km/s}$, and the ion distribution is initially drifting Maxwellian. For the isotropic reference case, the ion temperature is $T_i = 0.0003~\mathrm{eV}$, corresponding to $T_i/T_e = 0.001$.
The plate is initially kept at a positive bias, $\phi_{\mathrm{plate}} = 2\,\mathrm{V}$, such that the incoming ions encounter a repulsive electrostatic potential. This configuration favors ion compression ahead of the obstacle, thereby supporting the excitation of compressive precursor structures. Additional results illustrating the influence of different positive plate potentials on precursor formation are provided in Appendix~\ref{app: different_potential}. Moreover, the validation demonstrating that the PIC-generated upstream structures satisfy the characteristic properties of KdV solitons is provided in Appendix~\ref{app: validation_PIC}.
\subsection{Positive plate bias ($\phi_{\rm p}=+2V$) and cold isotropic ions ($T_i/T_e = 0.001$)}
\label{subsec:smallTi}
\begin{figure}
\centering
\includegraphics[width=1.0\linewidth, height=6cm]{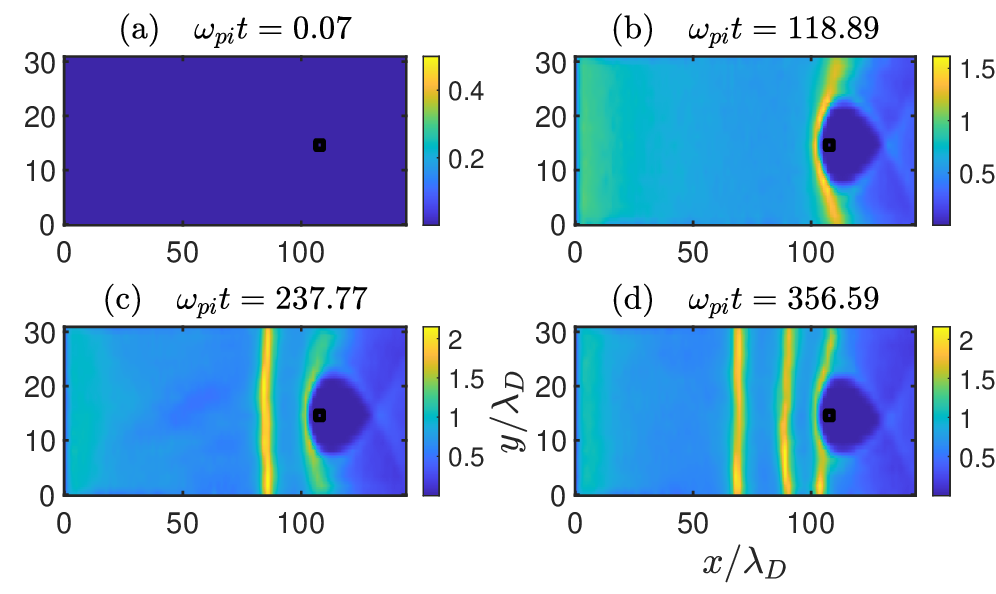}
\caption{Temporal evolution of the normalized ion density over a positively biased obstacle, demonstrating the emergence and evolution of upstream solitary excitations for $T_i/T_e = 0.001$. Snapshots are shown at (a) $\omega_{pi} t = 0.07$, (b) 118.89, (c) 237.77, and (d) 356.59.}
\label{fig:cold_density_main}
\end{figure}

\begin{figure}
\centering
\includegraphics[width=1.0\linewidth, height=6cm]{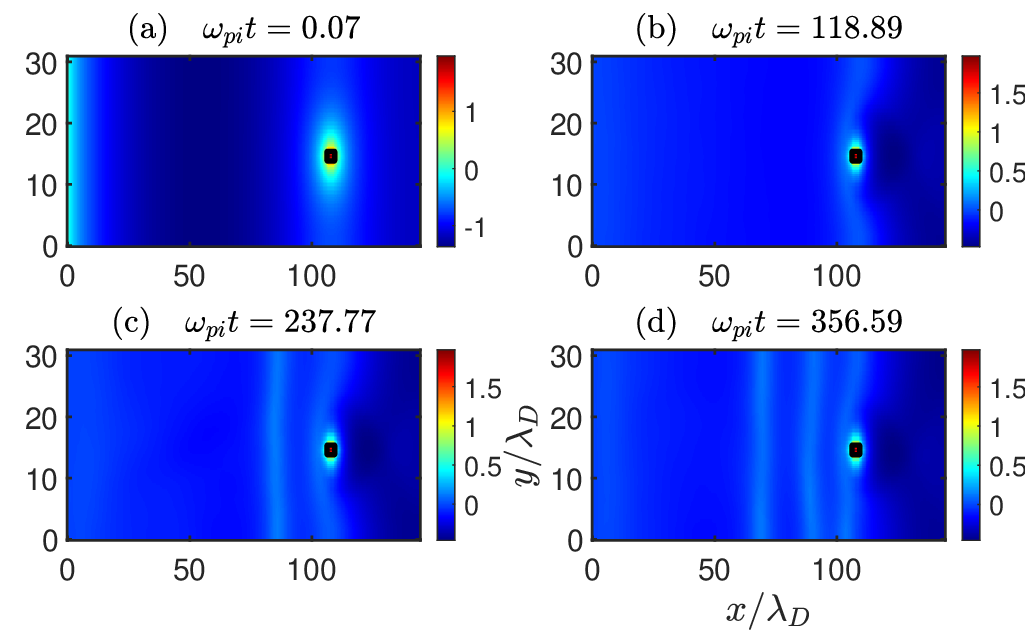}
\caption{Temporal evolution of the plasma potential over a positively biased obstacle, demonstrating the emergence and evolution of upstream solitary excitations for $T_i/T_e = 0.001$. Snapshots are shown at (a) $\omega_{pi} t = 0.07$, (b) 118.89, (c) 237.77, and (d) 356.59.}
\label{fig:cold_potential}
\end{figure}

We first consider the case of very cold isotropic ions, $T_i/T_e = 0.001$, with a positive plate potential $\phi_{\rm p}=+2V$. This temperature regime, as discussed in appendix~\ref{app:landau_damping}, is particularly favorable to the formation of solitary structures, since a larger fraction of the relevant mode spectrum experiences weak or negligible Landau damping \cite{resendiz2024ion,goldston2020introduction}. Consequently, the streaming ions can sustain a collective, and phase-coherent response to the electrostatic perturbation introduced by the obstacle. 
The density and potential evolution of the plasma interacting with the charged obstacle are presented in Fig.~\ref{fig:cold_density_main} and~\ref{fig:cold_potential}, respectively.
In the early phase of the density and potential profiles, $t\omega_{pi}=0.07$, shown in Fig.~\ref{fig:cold_density_main}(a) and~\ref{fig:cold_potential}(a), respectively. 
By $t\omega_{pi}=118.89$, a pronounced density compression develops upstream of the plate [Fig.~\ref{fig:cold_density_main}(b)] as ions accumulate ahead of the repulsive potential barrier. Owing to the finite extent of the obstacle, the compression initially exhibits a curved wavefront. The associated potential perturbation in Fig.~\ref{fig:cold_potential}(b) confirms the growth of the upstream disturbance.
At $t\omega_{pi}=237.77$, the compression steepens into a detached solitary pulse propagating upstream [Fig.~\ref{fig:cold_density_main}(c)]. The corresponding potential hump in Fig.~\ref{fig:cold_potential}(c) remains coupled to the density enhancement.
By $t\omega_{pi}=356.59$, multiple precursor solitons are emitted and propagate upstream [Fig.~\ref{fig:cold_density_main}(d)]. While the initial density compression near the obstacle exhibits a curved wavefront that reflects the local two-dimensional geometry of the plate potential, the emitted solitons extend over transverse scales much larger than the obstacle size and therefore appear nearly planar during propagation. The corresponding potential structures remain closely coupled to the density pulses [Fig.~\ref{fig:cold_potential}(d)]. 
%

The physical mechanism underlying the evolution of the upstream structures can be understood as follows. The positively biased plate repels the incoming ions, creating a localized compression region upstream. Because the plasma flow is weakly supersonic, the system lies near the threshold where finite-amplitude perturbations steepen into nonlinear wave trains. Simultaneously, the extremely low ion temperature strongly reduces Landau damping, preventing rapid dissipation of the injected energy. Consequently, the perturbation remains phase coherent, steepens, and separates into multiple precursor-like solitary pulses. These conditions are precisely those under which ion-acoustic precursor solitons are expected to form~\cite{resendiz2024ion}.
Conclusively, Figs.~\ref{fig:cold_density_main} and \ref{fig:cold_potential} together demonstrate that very cold ions with a positively charged obstacle provide an optimal environment for the generation of upstream nonlinear precursor structures. 
%
%
%
\subsection{Positive plate bias ($\phi_{\rm p}= 2V$) and temperature ratio ($T_i/T_e = 0.1$)}
\label{subsec:finiteTi_anisotropy}
We next examine the positive-bias case at a higher ion temperature ratio, $T_i/T_e = 0.1$, in order to isolate the combined effects of thermal spread and velocity-space anisotropy on precursor formation. In contrast to the cold-ion regime, the isotropic plasma case ($v_{\mathrm{th},iy}=v_{\mathrm{th},ix}$) where $v_{\mathrm{th},iy}$ and $v_{\mathrm{th},ix}$ are the ion thermal velocities perpendicular to the drift direction (along the $x$-axis) and along the drift direction, respectively, at $T_i/T_e=0.1$ does not support a well-defined upstream precursor wave train. It is consistent with the analysis in Appendix~\ref{app:landau_damping}.
As shown in Fig.~\ref{fig:finiteTi_density}, no distinct solitary pulse emerges even at longer times. This indicates that, at this temperature conditions, the obstacle-induced compression does not radiate in the form of solitary structures.
%
This behavior is consistent with previous kinetic studies \cite{resendiz2024ion,sam2025self} and results presented in Appendix~\ref{app:landau_damping} of ion-acoustic precursor formation, which report that when the ion temperature approaches one-tenth of the electron temperature, ion-acoustic disturbances experience appreciable Landau damping~\cite{resendiz2024ion}. In that regime, only very long-wavelength perturbations can propagate efficiently, while shorter-scale compressive structures (generated by a few Debye-scale debris and plasma interactions) are rapidly attenuated before steepening into solitary pulses. Since precursor excitations generated near an obstacle typically originate on Debye-length scales, the damping at $T_i/T_e \sim 0.1$ strongly suppresses their nonlinear development. We therefore treat this isotropic case as a Landau-damped reference state and subsequently examine whether introducing velocity-space anisotropy can reduce the damping to recover organized precursors.
%
%
\begin{figure}
\centering
\includegraphics[width=1.0\linewidth, height=6cm]{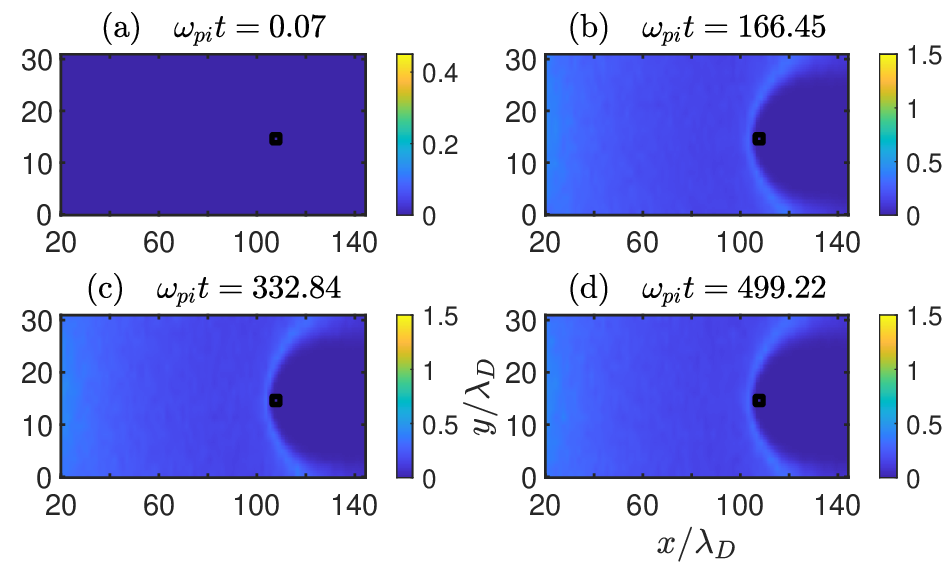}
\caption{Temporal evolution of the ion charge density over a positively biased obstacle for the isotropic case $T_i/T_e = 0.1$. The snapshots at (a) $\omega_{pi} t = 0.07$, (b) 166.45, (c) 332.84, and (d) 499.22 show no clear upstream solitary precursor is formed.}
\label{fig:finiteTi_density}
\end{figure}
\subsection{Effect of ion-temperature anisotropy with $\phi_{\rm plate}= 2V$ and $T_i/T_e = 0.1$}
\label{subsec:anisotropy}
We now introduce anisotropy in the ion thermal velocity distribution while keeping all other parameters fixed. The anisotropy is quantified by the ratio $v_{th,iy}/v_{th, ix}$, where smaller values correspond to stronger suppression of transverse thermal motion. 
The results presented in the following subsections for weak ($v_{th,iy}/v_{th,ix} = 0.1$), moderate ($v_{th,iy}/v_{th,ix} = 0.05$), and stronger ($v_{th,iy}/v_{th,ix} = 0.01$) anisotropies. It essentially demonstrates that it can induce precursor formation even when the ion temperature is not extremely low.
%
%
%
\subsubsection{Weak anisotropy: $v_{th,iy}/v_{th,ix} = 0.1$}
\label{subsubsec:anisotropy_01}
\begin{figure}
\centering
\includegraphics[width=1.0\linewidth, height=6cm]{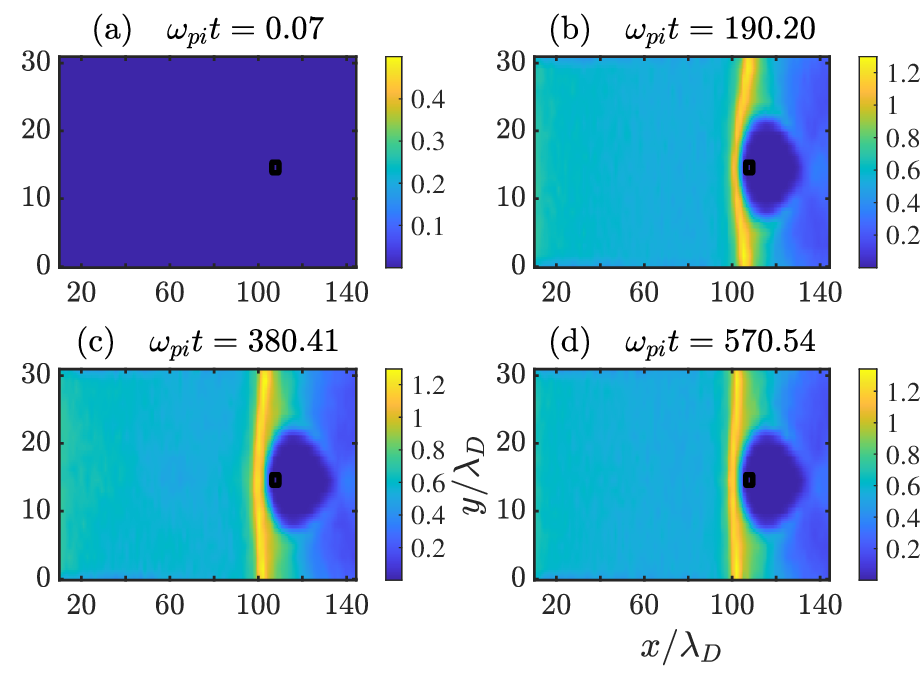}
\caption{Temporal evolution of the normalized ion density over a positively biased obstacle for the anisotropic case $v_{\mathrm{th},iy}/v_{\mathrm{th},ix}$=0.1 at $T_i/T_e=0.1$. The snapshots correspond to (a) $\omega_{pi} t = 0.07$, (b) $190.20$, (c) $380.41$, and (d) $570.54$.} 
\label{fig:anisotropy01_density}
\end{figure}

To assess the role of velocity-space anisotropy, we first consider the case $v_{th,iy}/v_{th,ix}=0.1$. Compared with the isotropic temperature case, the ion distribution is now more collimated, which enhances the upstream
compression in front of the positively biased plate. As seen in Fig.~\ref{fig:anisotropy01_density}, the initial state is shown in panel (a). At $t\omega_{pi}=190.20$ [Fig.~\ref{fig:anisotropy01_density}(b)], a clear density accumulation appears ahead of the obstacle, a feature that was absent in the isotropic case (Fig.4). This indicates a stronger compressive response due to anisotropic ion flow. At later times, $t\omega_{pi}=380.41$ and $570.54$ [Figs.~\ref{fig:anisotropy01_density}(c) and~(d)], the upstream density enhancement persists, but it does not break into a sharply localized precursor soliton train. Instead, the disturbance remains localized even at a longer time scale, suggesting that weak anisotropy strengthens the accumulation region, but it takes a sufficiently longer time to produce fully developed precursor structures.
\subsubsection{Moderate anisotropy: $v_{th,iy}/v_{th,ix} = 0.05$}
\label{subsubsec:anisotropy_005}
\begin{figure}
\centering
\includegraphics[width=1.0\linewidth, height=6cm]{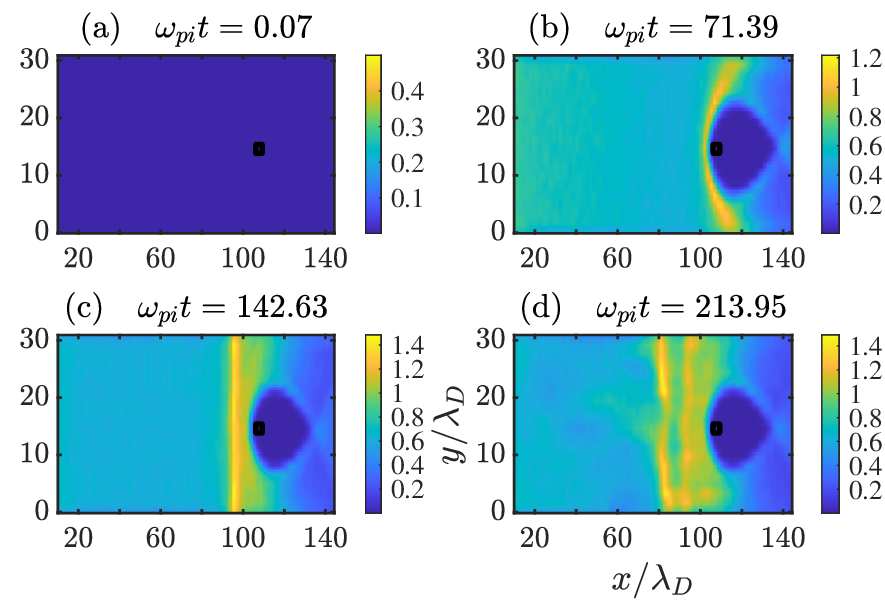}
\caption{Temporal evolution of the normalized ion density over a positively biased obstacle for the anisotropic case $v_{\mathrm{th},iy}/v_{\mathrm{th},ix}$=0.05 at $T_i/T_e=0.1$. The snapshots correspond to (a) $\omega_{pi} t=0.07$, (b) $71.39$, (c) $142.63$, and (d) $213.95$.} 
\label{fig:anisotropy005_density}
\end{figure}

A different response emerges when the anisotropy is increased to $v_{th,y}/v_{th,x}=0.05$. In this case, the ion distribution is more strongly collimated along the flow direction, thereby reducing transverse thermal spreading and enhancing nonlinear compression upstream of the positively biased plate. 
The density evolution is presented in Fig.~\ref{fig:anisotropy005_density}. By $t\omega_{pi}=71.39$
[Fig.~\ref{fig:anisotropy005_density}(b)], a localized density buildup is already visible ahead of the plate. At later times, $t\omega_{pi}=142.63$ and $213.95$ [Figs.~\ref{fig:anisotropy005_density}(c) and~(d)], this perturbation sharpens into a well-defined upstream solitary structure with clear precursor characteristics. The structure is substantially more localized compared to the $v_{th,y}/v_{th,x}=0.1$ case and remains coherent during propagation, indicating that nonlinear steepening dominates thermal smoothing. Thus, greater anisotropy enables the formation of precursor soliton-like excitations upstream of the obstacle.
\subsubsection{Strong anisotropy: $v_{th,iy}/v_{th,ix} = 0.01$}
\label{subsubsec:anisotropy_001}
\begin{figure}
\centering
\includegraphics[width=1.0\linewidth, height=6cm]{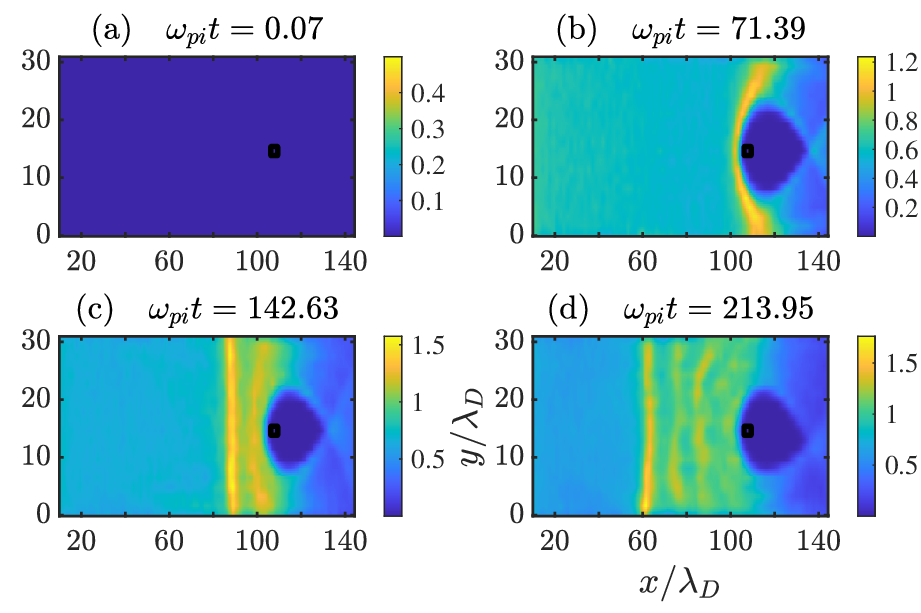}
\caption{Temporal evolution of the normalized ion density over a positively biased obstacle for the anisotropic case $v_{\mathrm{th},iy}/v_{\mathrm{th},ix}$=0.01 at $T_i/T_e=0.1$. The snapshots correspond to (a) $\omega_{pi} t=0.07$, (b) $71.39$, (c) $142.63$, and (d) $213.95$.}
\label{fig:anisotropy001_density}
\end{figure}

A sustained precursor formation is obtained for the larger anisotropy, i.e., considered here as $v_{th,y}/v_{th,x}=0.01$. In this case, the ion distribution is highly collimated along the flow direction, strongly suppressing transverse thermal spreading and allowing the obstacle-induced compression to remain coherent over time. 
In panel Fig.~\ref{fig:anisotropy001_density}(b), a pronounced density accumulation appears upstream of the positively biased plate, indicating the onset of nonlinear compression. At the next time stamp ($t\omega_{pi}$ = 142.63), this accumulated density reorganizes into a single propagating upstream structure [Fig.~\ref{fig:anisotropy001_density}(c)], demonstrating the emergence of a precursor solitary pulse. At a later time ($t\omega_{pi}$ = 213.95), multiple upstream structures are observed [Fig.~\ref{fig:anisotropy001_density}(d)], showing the development of a train of coherent solitary excitations. Among all the cases considered herein, a stronger anisotropy yields a more pronounced precursor excitation. 
%
%
These results suggest that reducing the perpendicular ion thermal spread weakens transverse smoothing and enhances nonlinear steepening ahead of the obstacle. The observed transition from diffuse compression to coherent precursor emission refers to the competition between transverse thermal dispersion and nonlinear ion compression in the streaming plasma. 
These results are also relevant to laboratory dusty-plasma experiments involving charged wires or embedded obstacles. In most existing experiments, precursor structures have been reported for negatively biased wires or obstacles interacting with a flowing cloud of negatively charged dust particles~\cite{bandyopadhyay2022driven}. Such configurations are physically analogous to the positively biased obstacle considered in the present work, in which a positively charged object interacts with a streaming background of positive ions. In both cases, the obstacle repels the incident charged species, thereby creating the upstream compression needed for precursor formation.
\subsection{Negative plate potential($\phi_p = -1.5V$)}
\label{subsec:negative_potential}
We next examine the case of a negatively biased obstacle. Unlike the positively biased configuration, a negative plate attracts ions toward itself rather than repelling them upstream. As a result, the localized upstream compression required for precursor excitation is not established under the present conditions. For $v_{\rm drift}=7~{\rm km/s}$ and $v_{th,y}/v_{th,x}=1.0$ , clearly no upstream precursor structure is observed. Instead, the density response is dominated by a downstream wake-like oscillatory pattern, indicating that the sign of the obstacle potential strongly influences whether the nonlinear plasma response evolves into a precursor train or a wake.
%
%
%
\begin{figure}[H]
\centering
\includegraphics[width=1.0\linewidth, height=6cm]{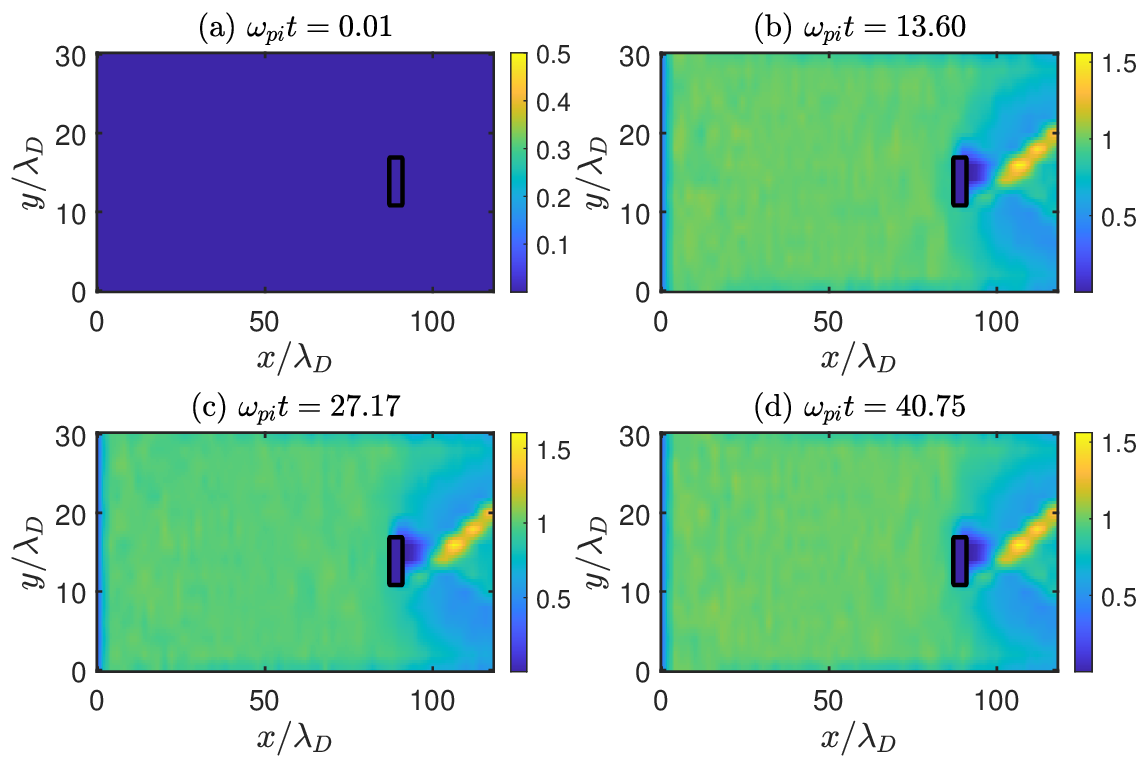}
\caption{Temporal evolution of the normalized ion density over a negatively biased obstacle for the isotropic case ($v_{\mathrm{th},iy}/v_{\mathrm{th},ix}$=1.0) and $T_i/T_e=0.001$. The snapshots correspond to (a) $\omega_{pi} t=0.01$, (b) $13.60$, (c) $27.17$, and (d) $40.75$.} 
\label{fig:negative_density}
\end{figure}
At the initial time, $\omega_{pi} t=0.01$, the perturbation remains weak [Fig.~\ref{fig:negative_density}(a)]. By $\omega_{pi} t=13.60$, the density begins to reorganize around the negatively biased plate. At later times, $\omega_{pi} t=27.17$ and $40.75$, a pronounced oscillatory wake develops in the downstream direction [Figs.~\ref{fig:negative_density}(b)--\ref{fig:negative_density}(d)], while no upstream solitary structure is formed. Thus, although the negative bias substantially modifies the ion flow, it does not generate the upstream compression necessary for precursor soliton formation in the present setup.
\section{Conclusion}
\label{sec:results_summary}
The numerical results presented in this work demonstrate that the excitation of upstream precursor structures in a streaming plasma is governed by a delicate interplay among ion temperature, ion-temperature anisotropy, flow coherence, and the polarity of the obstacle potential. The present study shows that precursor solitons are not a natural outcome of plasma flow past a charged obstacle; rather, they arise only when the obstacle-induced compression overcomes thermal spreading and kinetic damping, so that nonlinear steepening and localization are sustained.
%
%
Precursor formation is found to be most favorable under cold-ion conditions, where weak thermal effects preserve the coherence of the streaming ions and support robust upstream solitary excitations. Increasing the ion temperature enhances phase mixing and kinetic damping, leading to broader disturbances and a marked suppression of precursor structures. These results highlight the ion thermal state as a key factor governing nonlinear wave generation in streaming plasmas.

The most important outcome of the present work is that the finite
ion-temperature anisotropy might lead to the generation of sustained precursor structures. 
At a finite ion temperature, reducing the perpendicular thermal spread significantly improves the directional coherence of the flow. In the present study, anisotropy ratios such as $v_{th,iy}/v_{th,ix}=0.1$, $0.05$, and $0.01$ progressively enhance the localization of the upstream response. This confinement of thermal motion helps preserve the obstacle-induced compression and promotes the faster appearance of localized precursor structures. Stronger anisotropy leads to clearer and more persistent precursor excitations, demonstrating that anisotropy can partially compensate for the suppressive effects of finite ion temperature. The present results, therefore, identify anisotropy as an important kinetic control mechanism that must be accounted for in realistic plasma environments. These results suggest that reducing the perpendicular ion thermal spread weakens transverse smoothing and enhances nonlinear steepening ahead of the obstacle. The observed transition from diffusive compression to coherent precursor emission, therefore, reflects the
competition between transverse thermal dispersion and nonlinear ion compression in the streaming plasma. 

The polarity of the obstacle potential is also found to be decisive. A positively biased obstacle, here taken as $\phi_{\rm plate}= 2V$, generates the repulsive electrostatic compression necessary for precursor excitation, whereas a negatively biased obstacle does not support this mechanism and instead produces wake-dominated dynamics. 
This distinction confirms that precursor solitons depend not only on plasma parameters, but also on the nature of the external forcing.
%
%
%
For a negatively biased obstacle with absorbing boundaries, ions are largely removed from the flow by the obstacle, suppressing upstream compression and producing a wake-like response. In contrast, reflective or partially absorbing boundaries may allow a gradual buildup of density \cite{Mir}. Therefore, boundary conditions, ion reflection, and ion anisotropy must be further examined to assess the formation of stable nonlinear structures around negative-potential obstacles.

The broader implication of this work is that kinetic properties of the ion distribution play a fundamental role in determining the nonlinear plasma response to embedded charged objects. Since non-Maxwellian and anisotropic ion populations are common in both space and laboratory plasmas, the present results are relevant to a wide range of systems involving plasma flow past charged bodies, electrostatic barriers, and localized potential structures. This provides physics insight into obstacle-driven wave activity in ionospheric environments and related plasma experiments.

\section*{Conflicts of Interest}
The authors declare no conflicts of interest.
\section*{Data Availability Statement}
The data that support the findings of this study are available from the
corresponding author upon reasonable request.
\section*{Acknowledgments}
This work is supported by the Department of Space, Government of India. 
\appendix
\section{Numerical details}
\label{app:numerical details}
The computational domain consists of a two-dimensional Cartesian box ($L_x > L_y$), where $L_x$ and $L_y$ denote the directions parallel and perpendicular to the drifting plasma flow, respectively. The domain size is chosen sufficiently large to fully capture the evolution of the precursor solitons and their interaction with the plate. The domain is discretized using a uniform grid with suitably chosen resolution, such that the grid spacing satisfies $\Delta = 0.1\,\lambda_D$, where $\lambda_D=\sqrt{\varepsilon_0 T_e/(n_0 e^2)}$ is the  Debye length. The number of grid points in each direction is selected to ensure adequate numerical resolution of the plasma sheath, wake, and nonlinear wave dynamics. A plate, maintained at a fixed potential $\phi_p$, is embedded within the domain as a small rectangular obstacle of characteristic size comparable to the Debye length.
The plasma consists of singly charged ions ($M = 32\,\mathrm{amu}$) with background density $n_0 = 10^{12}\,\mathrm{m^{-3}}$ and electron temperature $T_e = 0.3\,\mathrm{eV}$. The ion temperature is varied in order to examine the influence of different temperature ratios $T_i/T_e$. Ions drift along the $x$-direction with velocity $v_{drift} = 1\,\mathrm{km\,s^{-1}}$, corresponding to a Mach number $v_{drift}/C_s \approx 1.1$, where $C_s=\sqrt{e(T_e+\gamma_iT_i)/M}$ . The ion thermal speed remains much smaller than the drift speed, $v_{th, i} < v_{drift}$.
%
The electrostatic potential is obtained from $\nabla^2\phi=-\frac{e}{\varepsilon_0}\left[n_i-n_0\exp\!\left(\frac{\phi-\phi_0}{T_e}\right)\right]$,
where $n_i$ is computed from particle deposition and the electron density follows a Boltzmann response. 
%
The governing equation is discretized using a standard five-point stencil and solved iteratively using a Gauss-Seidel scheme. As shown in Fig.~\ref{fig:Schematic}, boundary conditions are defined as $\phi=0$ at the left boundary ($x=0$), which also serves as the particle injection source, while homogeneous Neumann conditions $(\nabla \phi = 0)$ are imposed on the remaining outer boundaries. The internal plate is maintained at a fixed potential $\phi=\phi_p$. Particles are continuously injected from the left boundary and are removed from the simulation when they either exit the domain through the open boundaries or collide with the plate, where they are absorbed.
Ion macroparticles are advanced using a standard PIC cycle. Charge is deposited on the grid via cloud-in-cell interpolation, and electric fields are computed from finite differences of the potential and interpolated back to particle positions. 
%
%
At each time step, 
particles are injected near the left boundary with 
velocities drawn from a drifting Maxwellian, $f(v_x,v_y)\propto\exp\!\left[-\frac{(v_x-v_{drift})^2+v_y^2}{2v_{th,i}^2}\right]$.
The injection rate is chosen to maintain a steady inflow consistent with the flux $n_0 v_{drift}$.
The time step is $\Delta t = 0.1\,\Delta/v_{drift}$, ensuring that particles move a fraction of a grid cell per step and that $\omega_{pi}\Delta t \ll 1$. 
%
This setup self-consistently captures the interaction between a drifting plasma and a biased conducting obstacle, including density perturbations, wake formation, and ion-acoustic responses. 
\section{Validation of PIC-Generated Precursor Solitons}
\label{app: validation_PIC}
To assess whether the nonlinear structures reported in the main text indeed exhibit solitary characteristics, we perform a detailed validation using the PIC code. The simulations are carried out for plasma flow over a stationary obstacle
in the cold-ion-beam approximation, with the electron temperature fixed at $T_e = 0.3$ eV and $T_e \gg T_i$). 
This ordering suppresses ion Landau damping and supports the excitation and stable propagation of ion-acoustic solitons. 
In this regime, the nonlinear dynamics are well described by the fKdV equation, arising from the balance between nonlinear steepening and dispersive effects.

\begin{figure}[H]
\centering
\includegraphics[width=1.00\linewidth, height=5.9cm]{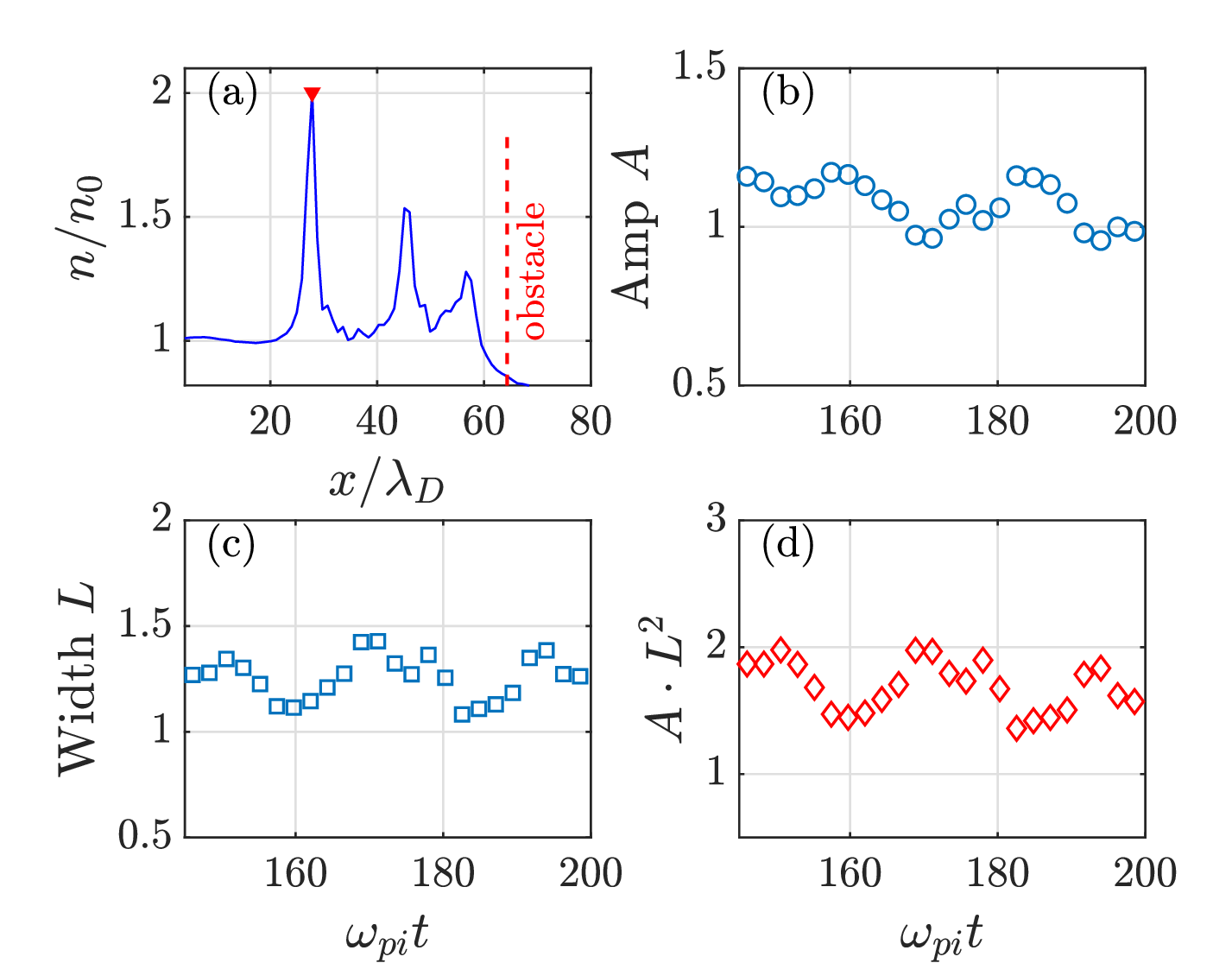}
\caption{Validation of precursor solitons in a cold ion beam plasma ($T_e=0.3$ eV). (a) Density profile showing upstream precursor formation. (b,c) Temporal evolution of soliton amplitude $A$ and width $L$. (d) Nearly constant $A L^2$ confirms KdV soliton behavior and validates the PIC simulation.}
\label{PIC_Kdv_Validation}
\end{figure}

Figure~\ref{PIC_Kdv_Validation}(a) shows a snapshot of the 1D spatial profile of the normalized ion density $n/n_0$ as a function of the normalized distance $x/\lambda_D$ at $t\omega_{pi} \approx 200$. To accurately resolve the small-scale precursor structures and their relatively narrow spatial widths, a finer spatial resolution is employed in the simulations. This 1D density profile is obtained from the two-dimensional data by averaging over the transverse ($ y$) direction, ensuring a statistically smooth representation of the longitudinal dynamics while preserving the essential physics of the propagating structures.
A train of localized density compressions is observed propagating upstream of the stationary obstacle located near $x/\lambda_D \approx 65$. These structures correspond to periodically emitted precursor solitons generated by the interaction between the flowing plasma and the obstacle. 
%
%
The leading precursor soliton (marked with a red inverted triangle in Fig.~\ref{PIC_Kdv_Validation} (a)) is tracked, and its amplitude (A) and width (L) are estimated by fitting the density peak with a $\mathrm{sech}^2$-shaped profile centered on the soliton.

The temporal evolution of the soliton amplitude ($A$) and width ($L$) is shown in Figures~\ref{PIC_Kdv_Validation}(b) and~(c) over the interval $\omega_{pi} t \in [145,200]$. 
This time window is chosen to track the fully developed soliton structure and to ensure that the structures remain well separated from subsequent emissions and boundary effects. 
As shown in Fig.~\ref{PIC_Kdv_Validation} (b), the amplitude remains nearly constant around $A \approx 1.0$, while the width, in Fig.~\ref{PIC_Kdv_Validation}, is maintained close to $L \approx 1.3 \lambda_D$, with only minor fluctuations.
A more rigorous test of the solitonic nature of these structures is shown in Figure~\ref{PIC_Kdv_Validation}(d), where the quantity $A L^2$ remains approximately constant over time. This confirms that the structures maintain the essential solitonic balance between nonlinearity and dispersion.
%
The close agreement between the PIC results and KdV invariants provides strong evidence that the precursor structures reported in the analysis are genuine solitons, thereby validating the capability of the developed code to accurately capture nonlinear soliton dynamics in plasma flow over a stationary obstacle.
%
%
\section{Temperature Dependence of Solitary-Wave Damping Relevant to the Present Parameters}
\label{app:landau_damping}
To examine the role of ion temperature and the possible influence of Landau damping on precursor formation, we performed an additional set of one-dimensional electrostatic PIC simulations for the propagation of an initially excited solitary disturbance under three different temperature ratios, namely $T_i/T_e=0.001$, $0.01$, and $0.1$, while keeping all other parameters fixed. The objective of this test is to isolate how the ion thermal spread in 1D conditions affects the survival, localization, and propagation of nonlinear solitary structures.
\begin{figure}[H]
\centering
\includegraphics[width=1.0\linewidth, height=8cm]{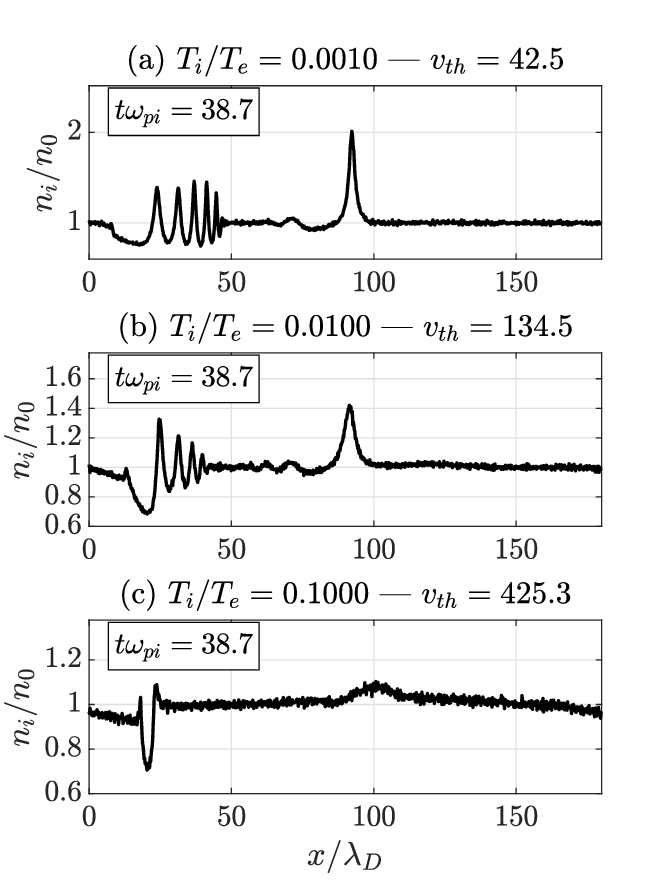}
\caption{Normalized ion density, $n_i/n_0$, plotted as a function of $x/\lambda_D$ at $t\omega_{pi}=38.7$ for three different temperature ratios: (a) $T_i/T_e=0.001$, (b) $T_i/T_e=0.01$, and (c) $T_i/T_e=0.1$. The solitary structure becomes progressively broader and its amplitude decreases with increasing $T_i/T_e$, indicating stronger damping at higher ion temperature.}
\label{fig:Landu Damping}
\end{figure}

The corresponding normalized ion density profiles, $n_i/n_0$, at $t\omega_{pi}=38.7$ are shown in Fig.~\ref{fig:Landu Damping}. A clear temperature dependence is observed. For the coldest case, $T_i/T_e=0.001$ [Fig.~\ref{fig:Landu Damping}(a)], the solitary pulse remains sharp, strongly localized, and retains a relatively large amplitude, indicating weak damping during propagation. For the intermediate case, $T_i/T_e=0.01$ [Fig.~\ref{fig:Landu Damping}(b)], a coherent solitary structure is still present, but its amplitude is moderately reduced and the profile becomes slightly broader. In contrast, for the hotter case, $T_i/T_e=0.1$ [Fig.~\ref{fig:Landu Damping}(c)], the solitary profile is substantially weakened in amplitude and significantly broadened in space. The pulse loses localization and coherence more rapidly than in the colder cases.

This trend is consistent with stronger kinetic damping at larger ion temperature. As $T_i/T_e$ increases, the ion velocity distribution becomes broader, thereby increasing the number of particles whose velocities are resonant with the propagating nonlinear disturbance. These resonant ions can exchange energy with the solitary structure through wave--particle interaction, removing energy from the coherent mode and suppressing a significant fraction of the spectral components required to maintain a steep and localized pulse. As a consequence, the disturbance spreads spatially and its peak amplitude decreases. In this sense, the hotter plasma case experiences stronger effective Landau damping, which inhibits the persistence of coherent nonlinear structures.
These observations are directly relevant to the precursor results presented in the main text. In the isotropic two-dimensional simulations with $T_i/T_e=0.1$, the same thermal broadening and damping tendency suppress the emergence of upstream precursor solitons, leaving only diffuse density accumulation near the obstacle. However, when velocity-space anisotropy is introduced, precursor structures reappear even at $T_i/T_e=0.1$. This suggests that anisotropy can partially mitigate the damping mechanism in two-dimensional systems by reducing the perpendicular thermal spread and increasing the directional coherence of the incoming ion stream. In effect, the ions become more collimated along the streaming direction, allowing stronger density accumulation ahead of the obstacle and enhancing the nonlinear steepening needed for precursor formation. 
\section{Effect of positive plate potential on the precursor structures}
\label{app: different_potential}
\begin{figure}
\centering
\includegraphics[width=1.0\linewidth,height=7cm]{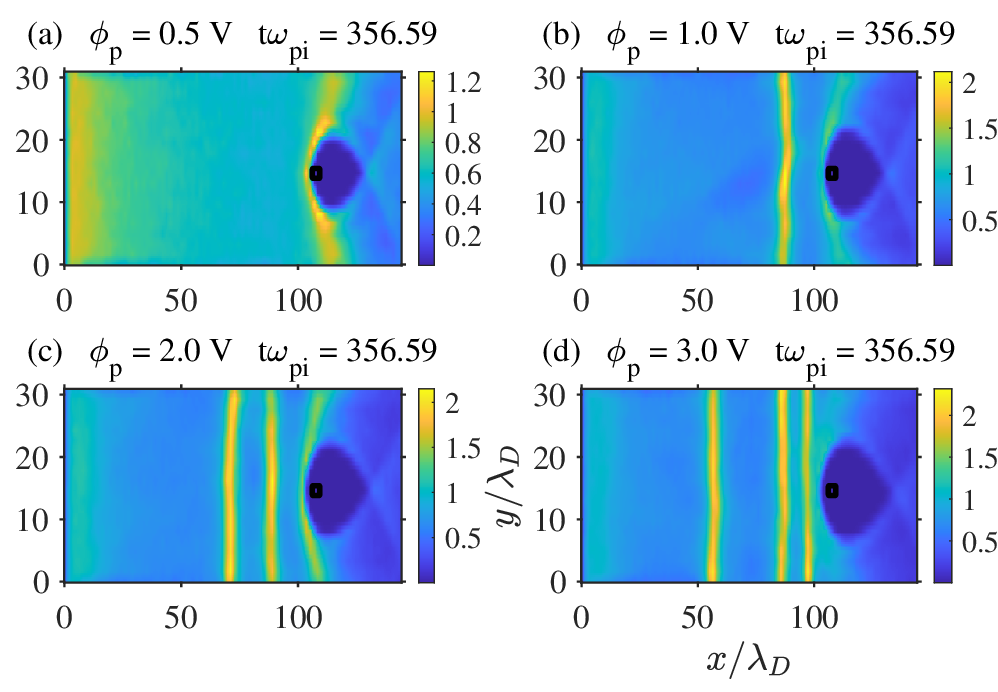}
\caption{Temporal snapshots of the normalized ion density for different positive plate potentials in the isotropic case $v_{th,y}/v_{th,x}=1.0$, with $T_i/T_e=0.001$, $v_{drift}$ = 1 $km/s$ taken at the fixed time $\omega_{pi} t=356.59$. Panels correspond to (a) $\phi_p=0.5~\mathrm{V}$, (b) $\phi_p=1.0~\mathrm{V}$, (c) $\phi_p=2.0~\mathrm{V}$, and (d) $\phi_p=3.0~\mathrm{V}$.}
\label{fig: Different_Potential}
\end{figure}

Figure~\ref{fig: Different_Potential} shows the ion density distribution at a fixed time $\omega_{pi} t = 356.59$ for different values of the positive plate potential, $\phi_p = 0.5 V,\; 1.0V,\; 2.0V,\; 3.0V$, while keeping the drift velocity and temperature conditions unchanged ($T_i/T_e=0.001$, $v_{th,iy}/v_{th,ix}=1.0$, $v_{drift}$ = 1 $km/s$). This allows us to isolate the role of the obstacle potential in controlling the excitation and evolution of upstream precursor structures.
A clear progression in the nonlinear response is observed as the plate potential is increased. For the $\phi_p = 0.5V$ [Fig.~\ref{fig: Different_Potential}(a)], the plasma exhibits only a localized density accumulation in front of the obstacle. The compression is relatively weak and does not evolve into a distinct propagating structure, indicating that the electrostatic forcing is insufficient to trigger nonlinear steepening.

As the potential is increased to $\phi_p = 1.0V$ [Fig.~\ref{fig: Different_Potential}(b)], the compression becomes stronger, and a single localized density peak emerges upstream of the plate. This marks the onset of precursor formation, where the balance between nonlinearity and dispersion begins to support a solitary-like structure.
For $\phi_p = 2.0V$ [Fig.~\ref{fig: Different_Potential}(c)], the upstream response becomes more pronounced, and multiple density peaks are observed. In this case, two distinct precursor solitons are formed, indicating that the stronger electrostatic repulsion enhances the nonlinear steepening and leads to the generation of a soliton train.
At the highest potential, $\phi_p = 3.0V$ [Fig.~\ref{fig: Different_Potential}(d)], the effect is further amplified. Three well-defined precursor structures are clearly visible in the upstream region. The increased plate potential strengthens the compression and accelerates the formation of successive solitons, resulting in a more extended and structured precursor train.
In addition to the increase in the number of solitons, the propagation speed of the structures also appears to increase with $\phi_p$. This behavior can be attributed to the stronger electrostatic potential gradient, which enhances the effective driving force on the ions and leads to faster-moving nonlinear structures.
Overall, these results demonstrate that the amplitude of the obstacle potential plays a crucial role in determining both the number and dynamics of precursor solitons. A threshold level of electrostatic forcing is required to initiate solitary-wave formation, and further increases in potential lead to a systematic growth in the number and speed of the upstream structures.
\FloatBarrier
\bibliographystyle{apsrev4-2}
\bibliography{paper.bib}
\end{document}